\documentclass[runningheads,a4paper]{llncs}

\usepackage{amssymb}
\usepackage{graphicx}

\begin{document}

\mainmatter  

\title{An artificial neural network application\\
on nuclear charge radii}

\titlerunning{An artificial neural network application on nuclear charge radii}

%
%
\author{Serkan Akkoyun, Tuncay Bayram, S. Okan Kara and Alper Sinan}
\authorrunning{Serkan Akkoyun, Tuncay Bayram, S. Okan Kara and Alper Sinan}

\institute{Department of Physics, Cumhuriyet University, Sivas, Turkey\\
Department of Physics, Sinop University, 57000 Sinop, Turkey\\
Department of Physics, Nigde University, Nigde, Turkey\\
Department of Statistics, Sinop University, 57000 Sinop, Turkey}
%
%

\toctitle{Nuclear charge radii} \tocauthor{S. Akkoyun et al.}
\maketitle

\begin{abstract}
The artificial neural networks (ANNs) have emerged with successful applications in nuclear physics as well as in many fields of science in recent years. In this paper, by using (ANNs), we have constructed a formula for the nuclear charge radii.  Statistical modeling of nuclear charge radii by using ANNs has been seen as to be successful. Also, the charge radii, binding energies and two-neutron separation energies of Sn isotopes have been calculated by implementing of the new formula in Hartree-Fock-Bogoliubov (HFB) calculations. The results of the study shows that the new formula is useful for describing nuclear charge radii.

\end{abstract}

\section{Introduction}

Nuclear charge radii and, more generally,
distributions of charge-density give direct information on the
Coulomb energy of nuclei. Because of this reason, it has gained
attention for the nuclear mass formulas in recent six
decades~\cite{Buchinger94}. As is well known, the nuclear charge
radii is a fundamental property of atomic
nuclei~\cite{Ring80,Greiner96}. It can be measured by various
methods based on the electromagnetic interaction between the nucleus
and electrons or muons. Widely used methods are measurements of
transition energies in muonic atoms, elastic electron scattering
experiments, $K_{\alpha}$X-ray and optical isotope shifts. A
detailed discussions on these techniques for measuring of the rms
(root-mean-square) charge radii of nuclei can be found
in~\cite{Angeli09}. The latest advances in experimental techniques
have provided accessibility to experimental nuclear charge of more
nuclei far from $\beta$-stability line.

The nuclear charge radii related with exotic phenomena such as skin
and halo has become a hot topic~\cite{Zhang02}. With its accuracy,
the study of the nuclear charge radii is very important for a better
understanding of not only the proton distribution in nuclei but also
the skin and halo. For these reasons, if one get a simple and
reliable formula for nuclear charge radii, it can be provides
information for exotic nuclei and the effective nucleon-nucleon
interaction. For these purpose, a number of nuclear charge formulae
have been proposed as to be based on $A^{1/3}$ and $Z^{1/3}$
dependence. An excellent comprehensive study on nuclear charge
formula obtained from fitting of experimental data can be found
in~\cite{Zhang02} and reference therein. However, in these studies
experimental data were limited. New experimental data over more than
800 nuclei have been updated, recently~\cite{Angeli09}. Using the
new data, the parameters of the nuclear charge formulae can be
revised. In the present work, we used feed-forward artificial neural
networks (ANNs) as a different approach for obtaining a reliable and
simple nuclear charge formula. As is well known, feed-forward ANNs
may be viewed as an universal non-linear function
approximator~\cite{Hornik}. In recent years, ANNs have been used in
many fields in nuclear physics as in the other fields, such as
determination of one and two proton separation
energies~\cite{Athanas1}, developing nuclear mass
systematics~\cite{Athanas2}, identification of impact parameter in
heavy-ion collisions~\cite{David,Bass,Haddad}, estimating beta
decay half-lives~\cite{Costiris} and obtaining potential energy
curves~\cite{Akkoyun}. The fundamental task of the ANNs is to give
outputs in consequence of the computation of the inputs. We have
obtained new mass dependent simple nuclear charge formula by using
all the data from ANN whose inputs were taken from the latest
experimental data.   The obtained new formula is presented and used
in HFB calculations for Sn nuclei.

The paper is organised as follows. In Section 2, ANNs formalism and
Hartree-Fock-Bogoliubov (HFB) method are given briefly. In Section
3, the results of the study and discussions are presented. Finally,
a summary is given in Section 4.

\section{Theoretical Framework}

\subsection{ANNs}
ANNs are mathematical models that mimic the human brain. They
consist of several neurons which are processing units. The neurons
are connected each other via adaptive synaptic
weights~\cite{Haykin}. The ANN is composed of three main layers.
The first layer corresponds to input layer, the intermediate layer
is called hidden layer and the last one is output layer. In this
study, one input layer with a neuron, one hidden layer with forty
four neurons (h=44) and one output layer with a neuron ANN was used.
The total adjustable weights was 2h=88. 

The input layer neuron receives the data from outside and the output
layer neuron gives the results. The data is transmitted via weighted
connections between the neurons. The tangent hyperbolic
(sigmoidlike) and linear functions were used for hidden and output
units respectively. It has been proven that one hidden layer and
sigmoidlike  activation function in this layer are sufficient to
approximate any continuous real function~\cite{Cybenko}. 

The use of ANN is as two-step process, training and test stages. In
this work, a back-propagation algorithm with
Levenberg-Marquardt~\cite{Levenberg,Marquardt} for the training of
the ANN was used. In the training stage, ANN modifies its weights
until an acceptable error level between desired and predicted
outputs is attained. The error function which measures this
difference was mean square error (MSE). After an acceptable error
level, the trained ANN is tested over the data of interest. For
further and general background for the ANN, the reader is referred
to~\cite{Haykin}.

\subsection{HFB Method}
In the HFB formalism, a two-body Hamiltonian of a system of fermions
can be interpreted in terms of a set of annihilation and creation
operators. The ground-state wave function is described as the
quasi-particle vacuum and the linear Bogoliubov transformation
provides connection between the quasiparticle operators and the
original particle operators. The basic building blocks of the HFB
method are the density matrix and the pairing tensor, and
expectation value of the HFB Hamiltonian could be expressed as an
energy functional (Details can be found in~\cite{Ring80,Stoitsov05}). In terms of Skyrme forces, the HFB energy has the form
of local energy density functional contains the sum of the mean
field and pairing energy densities. These fields can be calculated
in the coordinate space~\cite{Ring80,Stoitsov05}. In this work, HFB
equations have been solved by expanding quasi-particle wave
functions that conserve axial symmetry and parity on a harmonic
oscillator basis expressed in coordinate space proposed by Stoitsov
et al.~\cite{Stoitsov05}. For pairing, Lipkin-Nogami method is
implemented by performing the HFB calculations with an additional
term included in the HF Hamiltonian. To obtain ground-state properties of Sn
isotopes prescription of Ref.~\cite{Stoitsov05} was taken into account. A number of effective Skyrme forces can be found in literature for correct prediction of
the nuclear ground-state properties. In this study, the Skyrme force SLy4~\cite{Chabanat98} has been used.

\section{Results and Discussions}

\begin{figure}
\centering
\includegraphics[height=10cm]{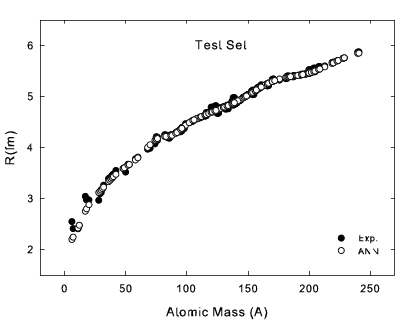}
\caption{The experimental (Exp.) and ANN test set predictions for nuclear radii for several isotopes ranging from A=6 to 240.}
\label{fig1}
\end{figure}

In this study, nuclear masses and experimental nuclear radii of $\sim$900 nuclei have been used in ANNs for fitting of the charge radii. The input of the ANN was atomic masses of the nuclei and the output was nuclear radii. The whole data was partitioned into two seperated sets, \%80 for the training stage and the rest for the test stage. After the training of the network, it was tested over the test set data which have never seen before by the network. As can be seen in the figure~\ref{fig1}, the ANN predictions agree exceptionally well with experimental values.

This obviously indicates that the test set ANNs have consistently generalized the training set fittings.
The MSE values were about 0.0027 and 0.0029 for the training and test stages respectively.
After testing of the trained network, the whole data from ANN outputs was used for employing least square
fitting. It should be noted that mass dependent form of nuclear charge radii given by the formula
$R_{c}=r_{0}A^{\beta}$ has been considered for fitting. The novelly obtained formula is given by
\begin{equation}
  R_{c} = 1.234 A^{0.28}
\end{equation}
where $R_{c}$ and $A$ is the rms charge radii and mass number of nuclei, respectively.
This formula are in good agrement with the experimental data. The root mean square deviation
between charge radii obtained from the formula and experimental ones is $\sigma=\sim1.4x10^{-2}$.
Beside, based on the formula we calculated the ground-state properties of nuclei such as binding
energy per nucleon, two-neutron separation energy and nuclear charge radii in HFB model for Sn isotopes.

\begin{figure}
\centering
\includegraphics[height=10cm]{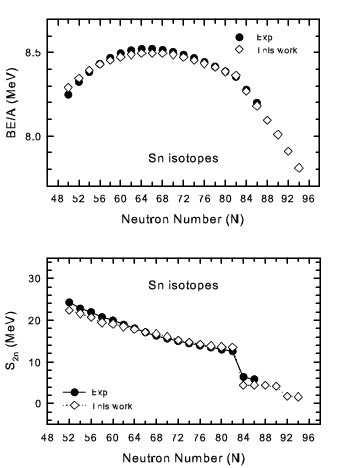}
\caption{Binding energies per nucleon (upper panel) and two-neutron separation energies (lower panel) for Sn Isotopes calculated with the spherical HFB code are compared with experimental data (Exp.) (from Ref.~\cite{Audi03}).}
\label{fig2}
\end{figure}

The calculated binding energies for even-even $^{100-144}$Sn isotopes within the HFB method with the Skyrme force SLy4 are shown in the upper panel of Figure~\ref{fig2}. Also, available experimental binding energies per nucleon taken from Ref.~\cite{Audi03} are shown for comparison. As can be seen in the upper panel of the Figure~\ref{fig2}, the predictions of HFB method are in good agrement with the experimental data. Maximum difference between the calculated and experimental values is only $\sim$0.04 MeV at neutron number $N=50$. In the lower panel of the Figure~\ref{fig2}, the calculated two-neutron separation energies of Sn isotopes and experimental ones are shown. Abrupt decrease of the two-neutron separation energy at neutron number $N=82$ indicates that this nuclei has a shell closure. Thus, shell effect is clearly visible in our calculations.

\begin{figure}
\centering
\includegraphics[height=5cm]{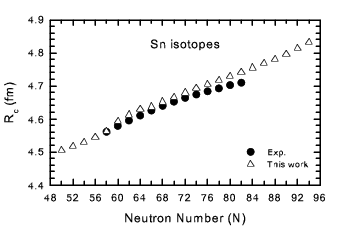}
\caption{The calculated rms nuclear charge radii within the HFB method with the Skyrme force SLy4 for Sn isotopes obtained by implementing the new formula in HFB code. Also, the experimental rms charge radii of Sn isotopes~\cite{Angeli09} are shown for comparison.}
\label{fig3}
\end{figure}

The calculated rms nuclear charge radii of $^{100-144}$Sn in HFB method by implementing the new formula and available experimental charge radii of Sn isotopes taken from Ref.~\cite{Angeli09} are shown in Figure~\ref{fig3}. As can be seen in the Figure~\ref{fig3}, the HFB method with the SLy4 parameters reproduced rms charge radii of Sn isotopes well. It should be noted, however, that ongoing from neutron number $N=58$ to neutron drip line, the difference between calculated and experimental charge radii for Sn isotopes is rising. However, maximum difference is only 0.032 fm at neutron number $N=82$.

\section{Summary}
ANNs have been applied on new experimental nuclear charge radii data to obtain simple charge radii formula based on mass dependence. The formula is founded as to be successful. By using the formula in HFB code, the ground-state properties of even-even $^{100-144}$Sn isotopes such as binding energies per nucleon, two-neutron separation energies and nuclear charge radii have been calculated. The results are founded as to be successful.

\end{document}